\documentclass[prb,aps,showpacs,groupedaddress,superscriptaddress,twocolumn]{revtex4-2}

\usepackage{natbib}
\usepackage[table]{xcolor}
\usepackage[colorlinks=true,linkcolor=blue,urlcolor=blue,citecolor=blue]{hyperref}

\usepackage{amsmath,amssymb}
\usepackage{float}
\usepackage{listings}
\usepackage{color}
\usepackage{graphicx}
\usepackage{nicefrac}
\usepackage{braket}
\usepackage{orcidlink}
\newcommand{\upperRomannumeral}[1]{\uppercase\expandafter{\romannumeral#1}}

\begin{document}

\title{Neural Network Quantum States for the Interacting Hofstadter Model with Higher Local Occupations and Long-Range Interactions}

\author{Fabian Döschl\orcidlink{0009-0005-5067-004X}}
\email{Fa.Doeschl@physik.uni-muenchen.de}
\affiliation{Ludwig-Maximilians-University Munich, Theresienstr. 37, Munich D-80333, Germany}
\affiliation{Munich Center for Quantum Science and Technology, Schellingstr. 4, Munich D-80799, Germany}

\author{Felix A. Palm\orcidlink{0000-0001-5774-5546}}
\affiliation{Ludwig-Maximilians-University Munich, Theresienstr. 37, Munich D-80333, Germany}
\affiliation{Munich Center for Quantum Science and Technology, Schellingstr. 4, Munich D-80799, Germany}
\affiliation{CENOLI, Universit\'e Libre de Bruxelles, CP 231, Campus Plaine, B-1050 Brussels, Belgium}

\author{Hannah Lange\orcidlink{0000-0002-0051-2087}}
\affiliation{Ludwig-Maximilians-University Munich, Theresienstr. 37, Munich D-80333, Germany}
\affiliation{Munich Center for Quantum Science and Technology, Schellingstr. 4, Munich D-80799, Germany}
\affiliation{Max-Planck-Institute for Quantum Optics, Hans-Kopfermann-Str.1, Garching D-85748, Germany}

\author{Fabian Grusdt\orcidlink{0000-0003-3531-8089}}
\affiliation{Ludwig-Maximilians-University Munich, Theresienstr. 37, Munich D-80333, Germany}
\affiliation{Munich Center for Quantum Science and Technology, Schellingstr. 4, Munich D-80799, Germany}

\author{Annabelle Bohrdt\orcidlink{0000-0002-3339-5200}}
\affiliation{Munich Center for Quantum Science and Technology, Schellingstr. 4, Munich D-80799, Germany}
\affiliation{University of Regensburg, Universitätsstr. 31, Regensburg D-93053, Germany}

\date{\today}

\begin{abstract}
Due to their immense representative power, neural network quantum states (NQS) have gained significant interest in current research. 
In recent advances in the field of NQS, it has been demonstrated that this approach can compete with state-of-the-art numerical techniques, making NQS a compelling alternative, in particular for the simulation of large, two-dimensional quantum systems. 
In this study, we show that recurrent neural network (RNN) wave functions can be employed to study systems relevant to current research in quantum many-body physics. 
Specifically, we employ a 2D tensorized gated RNN to explore the bosonic Hofstadter model with a variable local Hilbert space cut-off and long-range interactions. 
At first, we benchmark the RNN-NQS for the Hofstadter-Bose-Hubbard (HBH) Hamiltonian on a square lattice. 
We find that this method is, despite the complexity of the wave function, capable of efficiently identifying and representing most ground state properties.
Afterwards, we apply the method to an even more challenging model for current methods, namely the Hofstadter model with long-range interactions. 
This model describes Rydberg-dressed atoms on a lattice subject to a synthetic magnetic field. 
We study systems of size up to $12 \times 12$ sites and identify three different regimes by tuning the interaction range and the filling fraction $\nu$. In addition to phases known from the HBH model at short-ranged interaction, we observe bubble crystals and Wigner crystals for long-ranged interactions. 
Especially interesting is the evidence of a bubble crystal phase on a lattice, as this gives experiments a starting point for the search of clustered liquid phases, possibly hosting non-Abelian anyon excitations.
In our work we show that NQS are an efficient and reliable simulation method for quantum systems, which are the subject of current research. In particular, we demonstrate the ability of this method to simulate challenging systems with long-range interactions.
\end{abstract}

\maketitle
\section{Introduction\label{introduction}}

During the last decades, many exotic states of matter have been studied, a prominent example being fractional quantum Hall~(FQH) states.
Non-Abelian FQH states are of particular interest, due to their potential application in topological quantum computing \cite{Kitaev_2003,Nayak_2008}.
While historically such states have been first realized in solid-state experiments, cold atom quantum simulation provides a promising experimental platform allowing for a high degree of control, as well as interferometric approaches to braiding of anyons~\cite{Grusdt2016}.
In this cold atom setting, experiments have realized Laughlin states of a few bosons in rotating traps~\cite{gemelke2010rotating,Lunt2024} and in optical lattices~\cite{L_onard_2023}.
Building on this, proposals to grow larger states have been developed as well~\cite{Wang2023, Palm2024}.

The lattice systems are well-described by the Hofstadter-Bose-Hubbard model, which is known to host the (Abelian) Laughlin state~\cite{Laughlin_1983, Moller_2015, harper2015a, Raciunas_2018, Gra__2018, Macaluso_2020, Wang2023, Palm2024, FQHESoerensen, Hafezi_2007, Moller2009, Palm_2022, Repellin_2020, Dong_2018, He_2017, Hugel_2017, Palmer_2008} as well as the non-Abelian Pfaffian state~\cite{Moore1991,Sterdyniak_2012, Palm_2021, Boesl2022}.
However, numerical simulations of such systems remain challenging due to the intrinsic complexity of the system as well as the exponential growth of the Hilbert space in larger systems.
Typical numerical methods to tackle interacting many-body problems often suffer from the sign problem -- like quantum Monte Carlo (QMC) methods ~\cite{Foulkes_2001,Becca_Sorella_2017,Pan_2024} -- or are limited by the area law of entanglement -- like density matrix renormalization group (DMRG)~\cite{White1992} simulations based on matrix product states (MPS)~\cite{Schollw_ck_2011} --, which can become an issue when applied to extended two-dimensional systems.

While the existing literature mainly considered on-site Hubbard interactions, many interesting physical systems are governed by long-range interactions such as Coulomb interactions for electrons or van der Waals interactions in molecules or atoms.
As a prominent example, Rydberg atoms exhibit such long-range $\nicefrac{1}{r^6}$ van~der~Waals interactions. In the presence of a far off-resonant laser, the interaction length can be controlled, which can reveal intriguing physical properties~\cite{Macr__2014, Barbier_2022, Kazemi_2023}.
Such platforms offer strong and controllable atomic interactions with a large bandwidth of energy levels, which makes them an auspicious tool to realize correlated states of matter~\cite{Barredo_2018,Adams_2020, Browaeys_2020,Weber_2022}.
In the context of FQH systems, these Rydberg-dressed atoms are of special interest as they could potentially realize exotic clustered states~\cite{Grusdt_2013}.
The resulting cluster liquids are of particular interest, as they might realize the non-Abelian Pfaffian state~\cite{Moore1991} as well as even more exotic states.
Numerically, however, the long-range tails in larger systems are notoriously hard to simulate with existing methods.
In particular, DMRG with its spatially local optimization technique struggles with such interactions.
This raises the question of how these limitations can be overcome.

Here we show that neural network quantum states (NQS) along with variational Monte Carlo methods provide a promising framework to simulate long-range interacting many-body systems.
Such states represent a many-body wave function based on an artificial neural network, as first proposed in a seminal paper by Carleo~and~Troyer in 2017~\cite{Carleo_2017Solving}.
Since then, scientists have worked out many strengths and weaknesses of this ansatz~\cite{AnnealingHibat_Allah, Carrasquilla_2017, Carrasquillatutorial, EmpiricalGRU,Choo_2019}:
In recent publications, NQS have been successfully implemented for bosonic systems~\cite{BHLadder_even_2022, Saito_2017, Saito_2018, McBrian_2019, Vargas_Calder_n_2020, denis2024accurate, pei2024}.
In addition, it has been shown to be possible to capture topological phases in general with NQS~\cite{hibatallah2023investigating, Glasser_2018, Kaubruegger_2018, Li2021}.
A promising ansatz for efficient and accurate network architectures is given by using autoregressive networks, such as recurrent neural networks (RNNs), as they allow to sample directly from the wave function~\cite{Hibat_Allah_2020,  AnnealingHibat_Allah,wu2023tensor, Morawetz_2021, lange2023neural}.

Here we want to go beyond these studies and use an autoregressive model for bosonic systems with higher local occupations, topological ordered fractional quantum Hall states, and long-range interactions.
In particular, we employ a two-dimensional (2D), tensorized, and gated RNN~\cite{AnnealingHibat_Allah, VariationalHibat, SequTensor} to simulate a lattice model with a perpendicular magnetic field and long-range interactions.
The purpose of this study is to show the reliability of RNN-NQS for complex systems, especially in combination with higher local occupations, and the outstanding ability to describe models with long-range interactions.
The modifications of the plain vanilla RNN are particularly well suited for two-dimensional systems, as the information is passed through the system in a 2D manner. Additionally, the tensorized and gated structure is supposed to reduce the loss of information, which is important for capturing long-range correlations.

This paper is structured as follows: In Section~\ref{NQSDetails} we briefly review neural quantum states before extending the existing perfect sampling scheme for RNNs to systems with higher local boson numbers.
Furthermore, we emphasize the possibility to study systems with long-range interactions.
In Section~\ref{HofstadterModelApplications} we present applications of our modified RNN to strongly interacting systems exhibiting topological phases.
We start by motivating the Hofstadter-Hubbard model as a well-studied lattice version of the FQH problem (Sec.~\ref{NoninteractingHofstadter}) and benchmark our method against exact diagonalization (ED) and DMRG simulations (Sec.~\ref{BenchmarkHubbardInteractions}).
Afterwards, in Section~\ref{secLongRangeInt}, we apply our method to Rydberg-dressed atoms with long-range interactions and, for the first time, propose a phase diagram of this system on a lattice.
Numerical details are discussed in Appendices~\ref{app:NQS}~(RNN) and~\ref{app:DMRG}~(DMRG).
Additional numerical data is presented in Appendices~\ref{app:Benchmark} and~\ref{app:AdditionalData}.

\begin{figure*}
\centering
\includegraphics{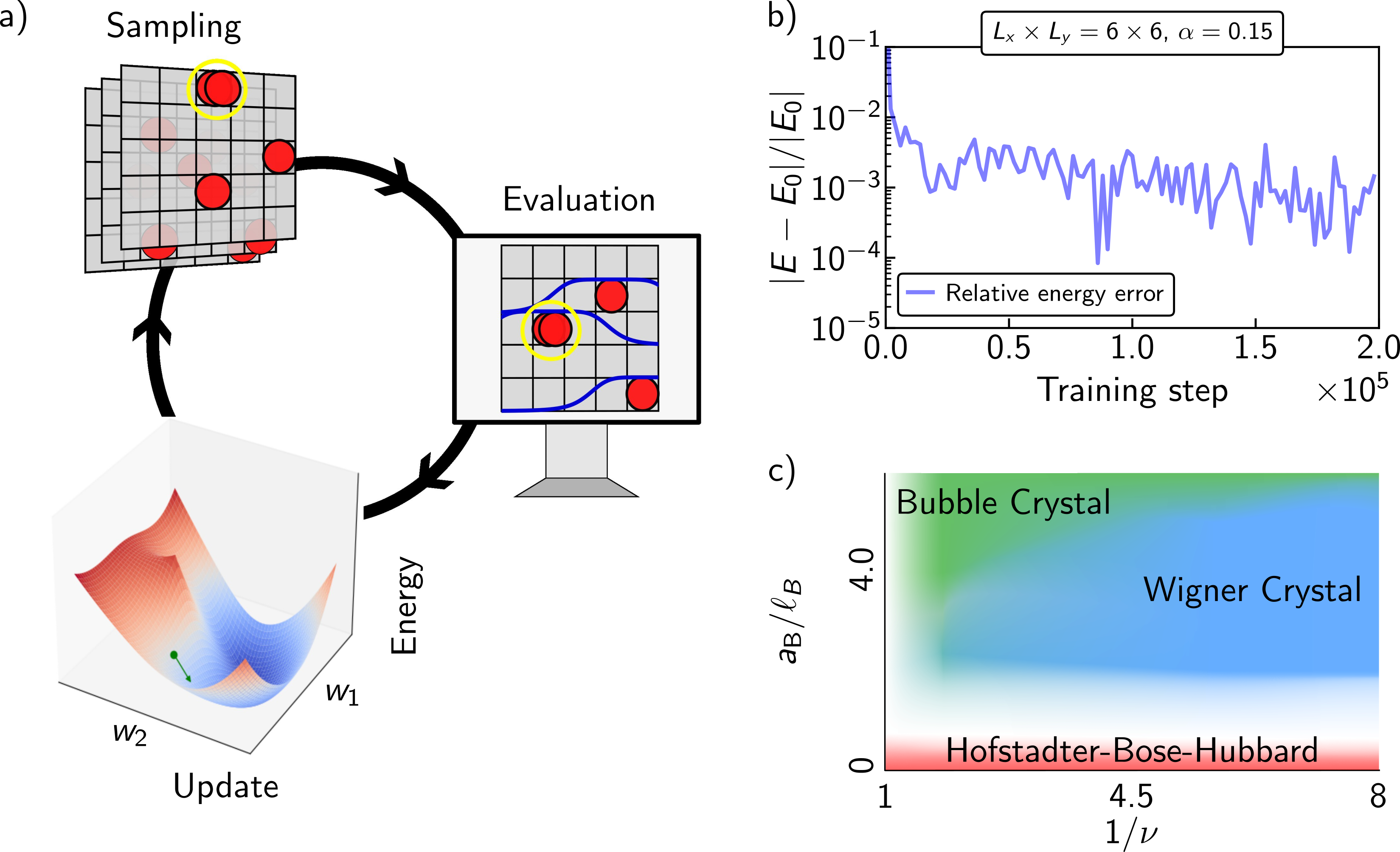}
\caption{A typical training procedure is displayed in panel a), b) shows the trainings progress of the RNN-NQS for $\alpha =0.15$ on the $6 \times 6$ lattice. Panel c) shows the phase diagram that we obtain with our NQS for a $12 \times 12$ lattice with long-range van der Waals interactions (see section~\ref{secLongRangeInt}). }
\label{NQS-Conv-Phase}
\end{figure*}

\section{Neural Network Quantum states}
\label{NQSDetails}
\subsection{NQS wave function}
We start by briefly reviewing the structure of neural quantum states ansatz. For further details we refer the reader to App.~\ref{app:NQS} and reviews~\cite{Hibat_Allah_2020, VariationalHibat, lange2023neural}.

The general idea is to use neural networks as a variational ansatz or a complex quantum state (see Fig.~\ref{NQS-Conv-Phase} a)), parametrized by the network's weights. To achieve this, the network's output with some parametrization is interpreted as a quantum mechanical wave function, typically in the Fock or spin basis. The parametrization of the output can be done in many different ways, which again depend on the network architecture~\cite{Saito_2017, McBrian_2019,chen2023efficient,oord2016pixel, TransformerZhang_2023}. Frequently used architectures include restricted Boltzmann machines~\cite{Carleo_2017Solving}, convolutional neural networks~\cite{roth2021group, Schmitt_2020, Choo_2019, chen2023efficient} and RNNs~\cite{oord2016pixel, Hibat_Allah_2020, VariationalHibat,wu2023tensor}.

In this study, we take advantage of the RNN's autoregressive property, which makes it possible to sample uncorrelated snapshots directly from the probability distribution. We define the RNN wave function as:
\begin{equation}
|\psi_\lambda\rangle = \sum_n \psi_\lambda(n) |n\rangle =\sum_n \exp(i \phi_\lambda(n))\sqrt{P_\lambda(n)} |n\rangle,
\label{Eq1}
\end{equation}
similar to Refs.~\cite{Hibat_Allah_2020, VariationalHibat,wu2023tensor, lange2023neural}. The phase $\phi_\lambda(n)$ and amplitude $P_\lambda(n)$ represent the network's output for one specific configuration  $n = \{n_{1}, n_{2},...\}$ in the Fock basis with the one-hot encoded particle number $n_{s}$ at site $s$~\cite{gori2023machine}. To improve the results with our model, we use a two-dimensional tensorized RNN (see Appendix~\ref{app:2D_RNN}) with one update gate~\cite{SequTensor, VariationalHibat, AnnealingHibat_Allah, lange2023neural}. This structure offers enhanced performance compared to the plain vanilla RNN cell~\cite{VariationalHibat}. In particular, the gate and the tensorized cell should help to capture long-range correlations, while the two-dimensional structure improves the accuracy of the RNN-NQS for two-dimensional systems. Note that the structure of the hidden state passing can be chosen depending on the boundary conditions used.

During the learning phase, the network parameters are optimized such that the energy is minimized. For this, we use the AdaBound optimizer~\cite{luo2019adaptive}, which offered the most reliable results compared to other tested state-of-the-art optimizers (see Appendix~\ref{comp6x6Sec}).

\subsection{Boson occupancy}
There are two major challenges when applying the RNN wave function to bosonic systems. The first is to control the total particle number. This can be done by adding a chemical potential, a constraint in the loss function, or by implementing a particle conservation formalism in the NQS that only allows samples from the desired subspace. We use the latter method, as we found it to work best.  

\begin{figure}
\centering
\includegraphics{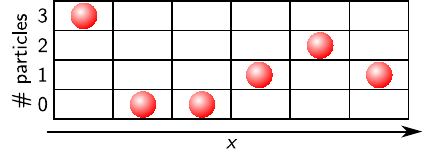}
\caption{Exemplary configuration, one-hot encoded, in one dimension with a maximum local particle number of \mbox{$N_\mathrm{max}=3$}.}
\label{Fig2BosonOccupation}
\end{figure}

The second major challenge is the, in principle, unrestricted boson occupation number per site. For spin~\cite{Choo_2019, AnnealingHibat_Allah, Wu_2019, roth2020iterative, Hibat_Allah_2020, Morawetz_2021} or fermionic systems~\cite{lange2023neural, Nomura2017, Inui_2021, Choo_2020}, the local Hilbert space dimension is naturally restricted, e.g. to $d_\mathrm{local}=2S+1$ for the frequently studied spin systems with spin $S$. When dealing with bosons, the possibility for a higher local occupation is desired. For numerical simulations, the local Hilbert space dimension is typically restricted to $d_\mathrm{local}=N_\mathrm{max}+1$, where $N_\mathrm{max}$ denotes a cut-off for the local maximal occupation number. This is illustrated in Figure~\ref{Fig2BosonOccupation}.

So far, bosonic NQS studies~\cite{BHLadder_even_2022, Saito_2018, McBrian_2019, Vargas_Calder_n_2020,denis2024accurate} tackled these challenges by using Metropolis Monte Carlo methods~\cite{MCSampling}, that can be restricted to draw candidate snapshots from the correct subspace.

Here, we extend these studies by applying the perfect sampling property of the RNN architecture for bosonic systems with a variable local Hilbert space cut-off. To this end, an algorithm must be implemented that adjusts the local particle number probabilities successively for all sites based on the total particle number and the local Hilbert space dimension. Ideally, this is done for all snapshots simultaneously, thus speeding up the sampling procedure tremendously. We use an algorithm similar to Ref.~\cite{barrett2022autoregressive}, which is explained in Appendix~\ref{AppU1}. 

\subsection{Long-range interactions}
One characteristic trait of nature is that most interactions have a long-range tail. Sometimes this tail is negligible, however, quite often not. Interactions like Coulomb interactions or van der Waals interactions decay proportional to a power law. In some numerical simulation methods like ED, such interactions can be taken into account, but such methods are limited by other factors like the exponentially increasing Hilbert space. Other state-of-the-art numerical methods, like DMRG~\cite{White1992,Schollw_ck_2011}, struggle with long-range interactions, especially when studying two- or three-dimensional systems.  

With NQS, in contrast, long-range interactions can be treated with relative ease. For typical density-density interactions, the distance of particles in each snapshot can be evaluated without much additional effort and the resulting interaction energy can simply be added to the diagonal part of the Hamiltonian in the $|n\rangle$-basis. Therefore, the additional computational cost is small.

\section{Application: Hofstadter model}
\label{HofstadterModelApplications}
Having introduced our main technical improvements, we now apply our algorithm to the paradigmatic Hofstadter model~\cite{Harper_1955} in the presence of interactions. We first show the applicability to bosonic systems in general by considering on-site Hubbard interactions in Sec.~\ref{BenchmarkHubbardInteractions}, before we turn to the case of long-range interactions in Sec.~\ref{secLongRangeInt}.

\subsection{From continuum to discrete lattice}
We consider a two-dimensional system subject to a magnetic field. The magnetic field introduces a length scale into the system, called the magnetic length $\ell_B \propto \nicefrac{1}{\sqrt{B}}$.
In general, this length scale might compete with other length scales of the system, resulting in the interesting behavior of many systems subjected to (strong) magnetic fields.
In the remainder of this work, we numerically investigate the interplay of the magnetic length with two examples of competing length scales.

First, we consider a square lattice with lattice constant $a$ and vary the magnetic length by tuning the flux per plaquette.
We furthermore add on-site Hubbard repulsion, resulting in interacting many-body states similar to previously studied fractional quantum Hall states on the lattice~\cite{Moller_2015, harper2015a, Raciunas_2018, Gra__2018, Macaluso_2020, Wang2023, Palm2024, FQHESoerensen, Hafezi_2007, Moller2009, Palm_2022, Repellin_2020, Dong_2018, He_2017, Hugel_2017, Palmer_2008, Andrews2021} as well as the non-Abelian Pfaffian state~\cite{Sterdyniak_2012, Palm_2021, Boesl2022}.
Other exotic states were found in numerical studies of lattice models with Chern numbers $|C|>1$~\cite{Andrews2018}.


Afterwards, we add another competing length scale by studying the effect of a finite Rydberg blockade radius $a_{\rm B}$.
Similar systems have previously been studied in the continuum~\cite{Grusdt_2013, PhysRevLett.104.195302, Cooper2005, Gra__2018, Mattioli_2013,Maghrebi_2015} and on a lattice model~\cite{Weber_2022}.
The competition of three length scales, $a$, $\ell_B$ and $a_{\rm B}$, is expected to lead to exotic phases of matter in a numerically challenging regime.

\subsection{Non-interacting model}
\label{NoninteractingHofstadter}
\begin{figure}
    \centering
    \includegraphics[scale=0.7]{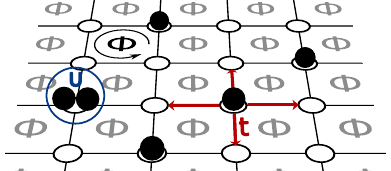}
    \caption{
        Illustration of the Hofstadter-Bose-Hubbard model. Bosons on a square lattice, with on site interaction $U$, hopping amplitude $t$, and magnetic Flux $\Phi$.
    }
    \label{HofstadterModelPNG}
\end{figure}

The non-interacting Hofstadter model~\cite{Hofstadter_1976} is a two-dimensional ($L_xL_y = A$), spinless, nearest neighbor tight-binding model for bosons with a hopping amplitude of $t$ and a perpendicular magnetic field. The constant magnetic field $B$ with flux $\Phi = B A$ is taken into account by Peierls phases in the hopping term. In the Landau gauge, the Peierls phase $\phi={2\pi\alpha x }$ is picked up during movement along the y-axis. Here, $\alpha$ is defined as the flux per plaquette $\alpha = \frac{\Phi}{(N_x-1)(N_y - 1)}$. The Hamiltonian describing this physical system reads (see Fig.~\ref{HofstadterModelPNG}):
\begin{equation}
\label{Hkin}
    \begin{split}
    \hat{H}_\mathrm{kin}(t,\alpha)=&-t \sum_{x}^{L_x-1}\sum_y^{L_y}(\hat{a}^\dagger_{x+1,y}\hat{a}^{\phantom{\dagger}}_{x,y} +h.c.) \\ &-t \sum_{x}^{L_x}\sum_y^{L_y-1}(\hat{a}^\dagger_{x,y+1}\hat{a}^{\phantom{\dagger}}_{x,y}e^{i2\pi\alpha x }+h.c.),
    \end{split}
\end{equation}
with the length $L_x$ and width $L_y$ of the system; the bosonic annihilation (creation) operators $\hat{a}^{(\dagger)}_{x,y}$. We further define the magnetic filling factor $\nu = N/N_{\phi}$, which is a ratio between the number of particles and the number of flux quanta. This well-studied model exhibits flat energy bands for small $\alpha$, similar to Landau levels known from the continuum. In the non-interacting limit, bosons will condense in the single-particle ground states.

\subsection{Hubbard interaction}
\label{BenchmarkHubbardInteractions}
In the following, we add on-site density-density interactions $U$ between bosons, which energetically penalize the condensation into a single mode. The resulting system is described by: 
\begin{equation}
\label{HBHeq}
    \begin{split}
    \hat{H}=\hat{H}_\mathrm{kin}(t,\alpha)+\frac{U}{2}\sum_{x}^{L_x}\sum_y^{L_y}\hat{n}_{x,y}(\hat{n}_{x,y}-1),
    \end{split}
\end{equation}
with the on-site density $\hat{n}_{x,y} = \hat{a}^{\dagger}_{x,y}\hat{a}^{\phantom{\dagger}}_{x,y}$. As in the case of the FQH effect, interactions lift the extensive ground state degeneracy and give rise to interesting phases~\cite{Kol1993}. For example, it was shown that the Hofstadter-Bose-Hubbard model harbors features of the Laughlin state at filling factor $\nu=\nicefrac{1}{2}$~\cite{Moller_2015, harper2015a, Raciunas_2018, Gra__2018, Macaluso_2020, Wang2023, Palm2024, FQHESoerensen, Hafezi_2007, Moller2009, Palm_2022, Repellin_2020, Dong_2018, He_2017, Hugel_2017, Palmer_2008}, and the Pfaffian state at $\nu=1$ ~\cite{Moore1991,Sterdyniak_2012, Palm_2021, Boesl2022}. 

In the following, we confirm the accuracy of our NQS approach by comparing it to a benchmark method. 
The compared observables are the energy and the expectation value of the interaction term.
The latter is defined as:
\begin{equation}
    C^{(2)}  = \sum_{x}^{L_x}\sum_y^{L_y} \frac{\langle\hat{n}_{x,y}(\hat{n}_{x,y}-1)\rangle}{N},
\end{equation}
where $N$ is the total particle number on the lattice. Note that the $ \nu = \nicefrac{1}{2}$-Laughlin is known to screen the Hubbard interaction, $C^{(2)}=0$~\cite{Palm_2021}.
Although this characteristic can be modified on the lattice, we still expect a strong suppression of $C^{(2)}$ to be a hallmark of this topologically ordered state.

\begin{table}[b]
\centering
  \begin{tabular}{c|c|c|c|c|c}
    System & hidden dim.& lr $\times 10^{-3}$ & samples & epochs $\times 10^3$& $d_\mathrm{local}$  \\
    \hline
     $6\times6$ & 100 & 0.5-0.05 & 200 & 200 & $3$\\
     $12\times12$ & 100 & 0.5-0.05 & 200 & $\lesssim 200$ & $3$
  \end{tabular}
  \caption{Hyperparameters used for the benchmark. They are defined in App.~\ref{app:DefHyper}). The number of parameters is 260806 in both cases. The Hilbert space dimension is 66045 for the $6\times6$ lattice and approximately $7.7\times 10^{20}$ for the $12\times12$ lattice. We used up to three different seeds: 1234, 4444, and 9999 for the calculations.}
  \label{Tab:TB}
\end{table}

\begin{figure*}
    \centering
    \includegraphics{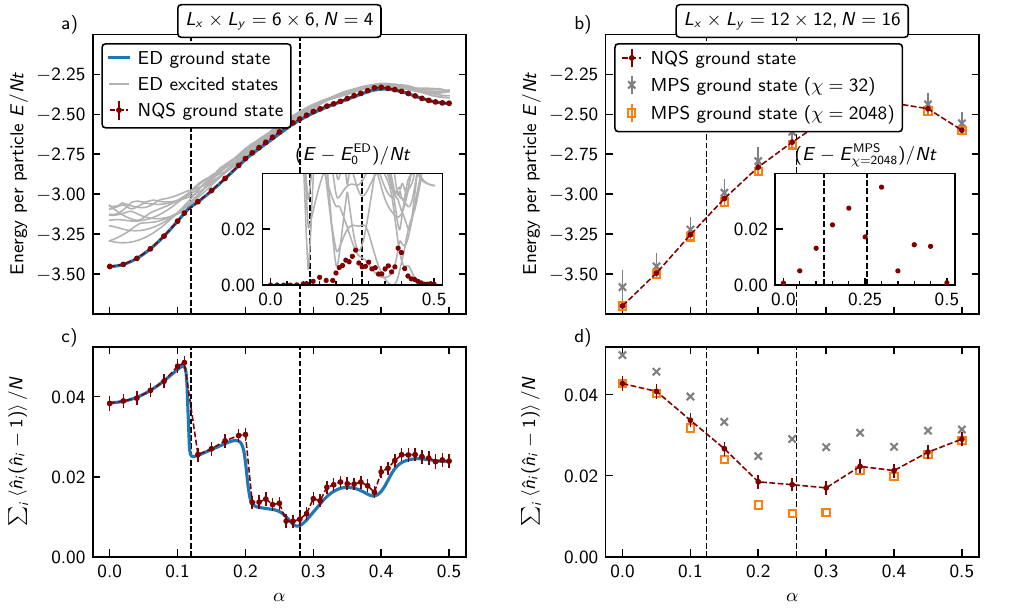}
    \caption{
        a,b)~Energy per particle and c,d)~Hubbard interaction energy of the ground state for two different system sizes and different numerical methods.
        a,c)~For a small system of $6\times6$ sites we find excellent agreement between the neural quantum state (NQS, red dots) and exact diagonalization results (ED, solid lines). 
        b,d)~For a larger system of $12\times12$ sites we find excellent agreement between NQS and DMRG ($\chi = 2048$) in the regime of low and high magnetic field. 
        In the intermediate regime between $\alpha = 0.2$ and $\alpha=0.3$ we find higher differences in the energy and the Hubbard interaction energy (see Appendix~\ref{app:topoligicalOrder}).
        Note that the NQS obtains better results than the parameter-wise comparable DMRG ($\chi=32$) simulation.
        For both systems, we considered a Hubbard repulsion of $U/t=4$ and a Hilbert space truncation to at most $N_{\rm max}=2$ bosons per site.
        The left vertical line represents the position $\nu_1 = N/(N-1)$, where features of the Pfaffian are to be expected. The right line corresponds to the filling fraction $\nu_{\nicefrac{1}{2}} = N/(2N-1)$ where the \nicefrac{1}{2} Laughlin state is expected.
        The error is the standard deviation of the mean, with an error propagation for the density-density correlation.
    }
    \label{FigHofEalpha}
\end{figure*}
%

To compare the NQS to exact diagonalization, we use a $6\times6$ square lattice with 4 particles. The results for the $12\times12$ square lattice with 16 particles are compared to DMRG. For both systems we use a local Hilbert space cut-off of $N_\mathrm{max}=2$, a Hubbard interaction $U=4t$ and vary the magnetic flux per plaquette $\alpha$.
The considered system has open boundaries, thus having a filling fraction of $\nu=\frac{N}{\alpha (L_x-1)(L_y-1)}$.

The comparison of important observables can be seen in Fig.~\ref{FigHofEalpha}.
The panels a) and c) show the results for the $6\times6$ lattice, the panels b) and d) display the results for the $12\times12$ lattice. Furthermore, we show the overlaps of the NQS with ED states for the $6\times6$ lattice in Fig.~\ref{FigHofOverlap}. We find a good agreement of the NQS results with the benchmark in the energy. Also the $C^{(2)}$ values of both methods match, except for a few challenging states on the large lattice. The used hyperparameters are given in Tab.~\ref{Tab:TB}. In the following, we discuss the results in detail.

\subsubsection{\texorpdfstring{$6\times 6$}{TEXT} square lattice}
\label{Benchmark6x6}
The energy plot in Fig.~\ref{FigHofEalpha}~a) shows that the NQS excellently manages to approximate the ground state energy in all cases for the $6\times 6$ system.
Further, the expectation value of the Hubbard interaction (Fig.~\ref{FigHofEalpha}~c)) barely deviates from the ED value.
This is quite remarkable, since the large number of low-lying excited states makes the training demanding. To evaluate our results, we distinguish between three different $\alpha$ regimes.

\begin{figure}
    \includegraphics{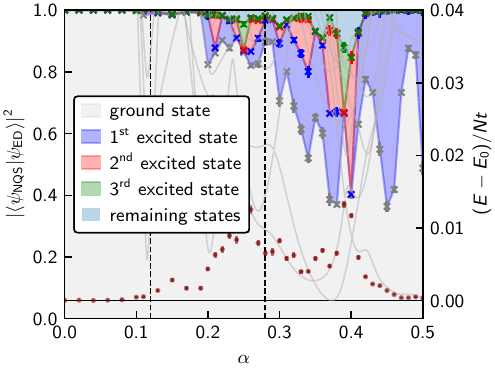}
    \caption{Overlap of the NQS with exact states from ED on the left axis. The right axis again shows the deviations from the ground state energy.
    }
    \label{FigHofOverlap}
\end{figure}

(\upperRomannumeral{1}) For $\alpha\lesssim 0.19$, especially in the topologically trivial phase for $\alpha< 0.11$, there are no deviations in the energy.
At $\alpha\approx 0.12, \nu \approx 1$ there is an energy gap closing, which makes it slightly more challenging for the RNN to learn the ground state.
For similar $\alpha$, earlier numerics found indications for a lattice Pfaffian state ~\cite{Sterdyniak_2012,Palm_2021,Boesl2022}, a topologically ordered state.
Between $ \alpha= 0.13$  and $\alpha= 0.19$ the NQS manages to get close to the ground state energy, but has a minimal offset.
When comparing the overlap with ED states, we find that the NQS almost perfectly captures the exact ground state (see Fig.~\ref{FigHofOverlap}).
Also, the results of $C^{(2)}$ are in excellent agreement with ED (Fig.~\ref{FigHofEalpha}~c)).

(\upperRomannumeral{2}) For $\alpha \in [0.2,0.33]$ and $\alpha \in [0.42,0.5]$ the training gets more challenging due to the increased complexity of the ground state and the energetically close excited states additionally complicate the training.
Nevertheless, the NQS achieves a relative energy error of less than $0.5\%$.
Especially around the topologically ordered $\nicefrac{1}{2}$ Laughlin at $\alpha\approx 0.28, \nu = 0.5$, we observe a ground state overlap of above $80\%$ (see Fig.~\ref{FigHofOverlap}). Substantial overlaps with the first and second excited states explain the deviations in the energy.

For $\alpha \in [0.42,0.5]$ the NQS finds a superposition of ground and first excited state. Note that these two states become quasi degenerate for $\alpha \rightarrow 0.5$.
Additionally, in this regime the $C^{(2)}$ value is with some minor deviations in good agreement with the exact results.
We conclude that the NQS manages to capture most characteristics of this regime, in particular also those of the topological ordered $\nicefrac{1}{2}$ Laughlin despite small deviations in the energy.

(\upperRomannumeral{3}) Between $\alpha= 0.34$ and $\alpha= 0.41$, the RNN energy is again very close to the ground state energy, with a relatively small error of less than $ 0.5\%$. The expectation value of the Hubbard interaction is also accurate, as we see all characteristic features in this region. Nonetheless, there are again some minor deviations in the $C^{(2)}$ value around $\alpha=0.4$. Despite the accurate prediction of the previous observables, we see larger contributions from multiple low-lying excited states in this regime (see Fig.~\ref{FigHofOverlap}). Especially the deviation in $C^{(2)}$ can be explained by a major contribution of the second excited state.

\subsubsection{\texorpdfstring{ $12 \times 12$}{TEXT} square lattice}
\label{Benchmark12x12}
When dealing with larger systems ($12 \times 12$), the NQS always reaches a low-lying state, with energies close to the DMRG ($\chi=2048$)~\footnote{For details on the DMRG simulations see~\ref{app:DMRG}} result, see Fig.~\ref{FigHofEalpha}~b,d). 
However, the topological order of some ground states is more challenging to capture than for the small system. It should be noted that the NQS only uses a fraction of the DMRG parameters (see table~\ref{Tab:TB}), whereas the results show only minor deviations in most cases. A DMRG simulation with a comparable number of variational parameters ($\chi=32$) yields significantly worse results than the NQS.

In the following, we will discuss the results in detail. We distinguish two different $\alpha$ regimes.

(\upperRomannumeral{1}) In the region $\alpha \lesssim 0.1$ and $\alpha \gtrsim 0.35$ we have energy deviations of less than $0.6\%$, which is comparable to the $6 \times 6$ case.
In this region, the NQS not only reaches a low energy value, but also the $C^{(2)}$ value is consistent with DMRG.
Note that we observe a lower variance in this regime compared to part (\upperRomannumeral{2}) ($\sigma^2_{(\mathrm{\upperRomannumeral{1}})}(E)/E^2 \lesssim \sigma^2_{(\mathrm{\upperRomannumeral{2}})}(E)/E^2 $, see Appendix~\ref{app:Benchmark}).

(\upperRomannumeral{2})~For $\alpha \in [0.1$  $0.3]$, the ground state is harder to capture for the RNN. Around unit filling, where the Pfaffian is a ground state candidate ($\alpha \approx 0.12$), the absolute energy error is marginally larger compared to part (\upperRomannumeral{1}) and the NQS $C^{(2)}$ value only differs slightly from the DMRG value. In the intermediate range ($0.2\lesssim\alpha \lesssim 0.30$), the relative energy error remains reasonable at approximately $1\% $.  However, there is a notable deviation from the $C^{(2)}$ benchmark value.
This regime appears to be challenging for the NQS, probably due to the topological order that the ground states exhibit in this regime.

\subsubsection{Comparison}
Scaling the system up from a $6\times6$ lattice to a $12\times12$ lattice exponentially increases the Hilbert space dimension from $H_\mathrm{dim}\approx 10^5$ to $H_\mathrm{dim} \approx 10^{21}$ and hence complicates the ground state search. Nevertheless, the NQS only slightly loses its energetic accuracy despite having the same number of parameters and learning for approximately the same number of epochs.
Furthermore, the NQS predicts the correct results for the expectation value of the interaction term for all magnetic fields for the $6 \times 6$ lattice. Here we showed that the ground state governs the properties of the trained NQS in almost all cases. Also, for the $12 \times 12$ lattice, we observe a good agreement between NQS and DMRG for low and high magnetic field. 
In the intermediate regime for the $12 \times 12$ lattice, where the system exhibits topological order, we observe higher deviations from the benchmark values.
Thus, we conclude that the NQS reaches an energy close to the ground state energy and captures, in most cases, its characteristics, even for very large systems. However, it gets more challenging when topological order is involved. In Appendix~\ref{app:topoligicalOrder} we discuss the difficulties of capturing topologically ordered ground states.

\subsection{Long-range interactions}
\label{secLongRangeInt}
In this section, the short-range interaction used in the previous section is replaced by long-range interactions. The new interaction length scale is an additional factor that influences the physical properties of the system. In combination with the magnetic length and the lattice constant this yields interesting physics. In the following we give a brief overview of previous studies that considered similar interactions in the continuum, see Sec.~\ref{prevStud}; thereafter, we show our results in section~\ref{phaseD}.

\subsubsection{Rydberg-dressing induced interactions}
\label{prevStud}
For experiments, Rydberg atoms are an outstanding tool due to their versatility which allows the realization of a great variety of quantum simulations. In particular, the high interaction strength between those excited atoms can be important for experiments. E.g., for fractional quantum Hall simulations, a large interaction strength is often desired. The natural Rydberg interaction scales with distance r as $\nicefrac{1}{r^6}$. However, in combination with a far-off resonant laser a highly controllable flat top potential can be realized:
\begin{equation}
U(r,a_\mathrm{B})=\frac{U_0}{a_\mathrm{B}^6 + r^6} a_\mathrm{B}^6.
\end{equation}
The blockade radius $a_\mathrm{B}$ depends on the frequency of the laser $\Omega\ll|\Delta|$ and its detuning parameter $\Delta$. For a more detailed description, we refer the reader to Ref.\cite{PhysRevLett.104.195302}.
These so-called Rydberg-dressed atoms allow us to study many systems of interest for current research in quantum many-body physics~\cite{Zeiher_2016, Hollerith_2022, Christakis_2023, Schine_2022}.
Using Rydberg-dressed atoms in an artificial gauge field is also a promising proposal for the realization of a controlled FQH system~\cite{Grusdt_2013, PhysRevLett.104.195302, Cooper2005, Gra__2018, Mattioli_2013,Maghrebi_2015}. In particular, a larger blockade radius was theoretically shown to favor clustered phases in the continuum and is, therefore, an interesting tool for the realization of potentially non-Abelian, clustered states such as the Pfaffian~\cite{Grusdt_2013}.

In the following, we investigate a two-dimensional lattice model with similar long-range interactions and propose a phase diagram that can be used as a starting point for further research. 
Hereby, we demonstrate that NQS can accurately approximate the ground state in phases that exhibit such interactions, which would constitute a significant challange to established numerical methods.

\subsubsection{Proposed phase diagram for systems on a finite-size lattice}
\label{phaseD}
\begin{figure}
    \centering
    \includegraphics{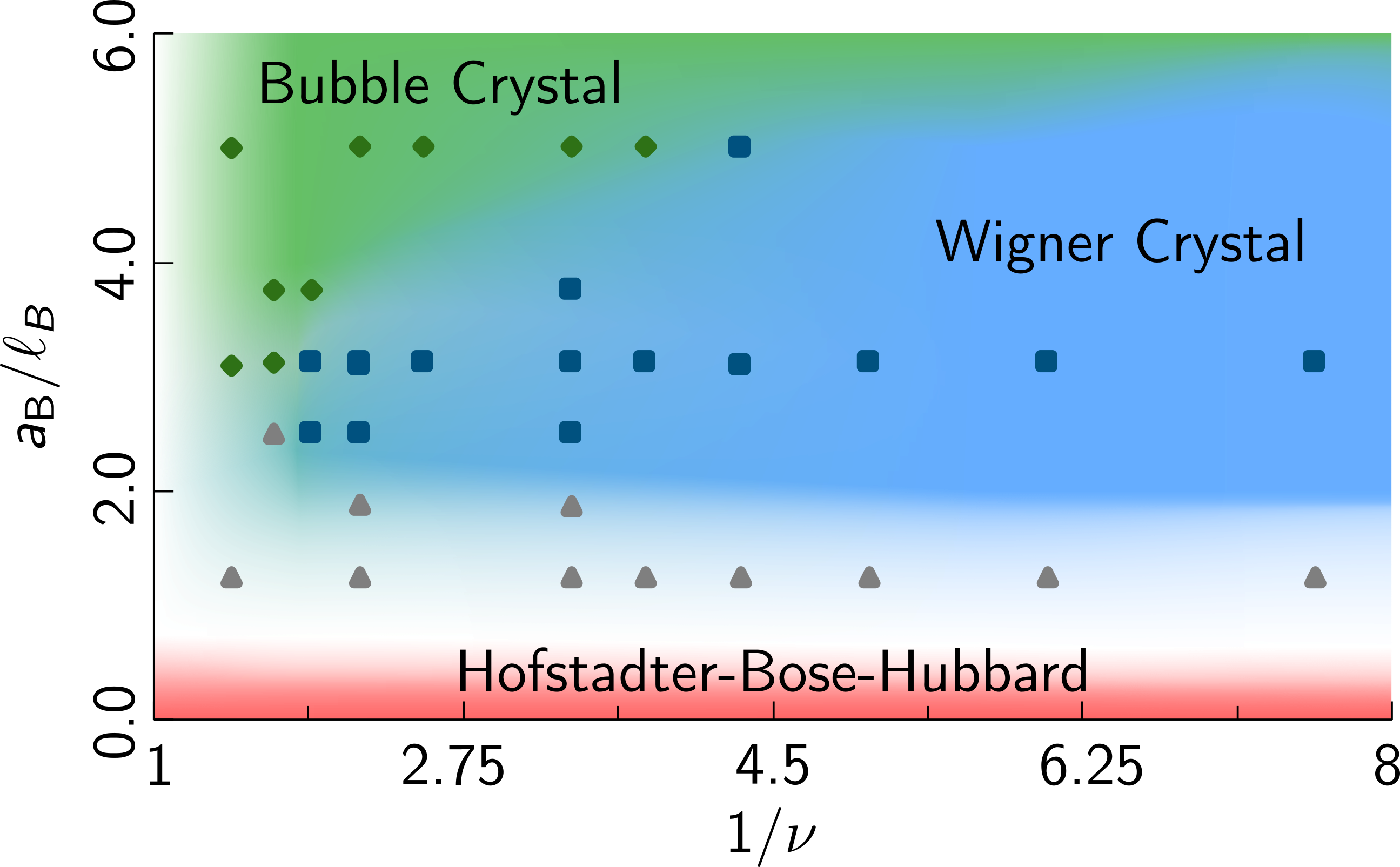}
    \caption{Finite-size phase diagram for the Hofstadter model with long-range Rydberg-dressed interactions obtained with NQS calculations. Marked in blue is a Wigner crystal phase, green is the bubble crystal phase. At $a_B \rightarrow 0$ we expect physics from on-site Hubbard-type interactions. To obtain this phase diagram we performed calculations at parameters corresponding to the symbols. The different symbols indicate which regime/phase the NQS ground state belongs to, while the background color is used as a guide to the eye. We used various numbers of particles with $N \in [4,21]$ for the $12 \times 12$ system.}
    \label{Phase}
\end{figure}

In this section we discuss the NQS-phase diagram for the simulation of Rydberg-dressed atoms on a finite-size lattice with a synthetic magnetic field. The Hamiltonian of the system reads:
\begin{equation}
\begin{split}
\hat{H}=&\hat{H}_\mathrm{kin}(t,\alpha)\\
&+\frac{U_0}{2} \sum_{x,x'}^{L_x}\sum_{y, y'}^{L_y}\frac{a_\mathrm{B}^6}{a_\mathrm{B}^6+r_{xy,x'y'}^6}\hat{n}_{x,y}(\hat{n}_{x', y'}-\delta_{x,x'}\delta_{y, y'}),
\end{split}
\end{equation}
with $\hat{H}_\mathrm{kin}(t,\alpha)$ defined in Eq.\eqref{Hkin}; the tunable blockade radius is denoted as $a_B$ and the particle number operator on lattice site $(x,y)$ is $\hat{n}_{x,y}$. We use a $12 \times 12$ lattice with open boundary conditions, an interaction strength of $U_0/t = 1$ and a magnetic flux per plaquette of $\alpha = 0.25$.

The phase diagram that we find for the considered finite-size system contains three different regimes, see Fig.~\ref{Phase}. The first is the Hofstadter-Bose-Hubbard (HBH) regime, which we obtain for $a_\mathrm{B}\rightarrow 0$. We know from the benchmark results in subsection~\ref{BenchmarkHubbardInteractions} that this regime can be represented by the RNN-NQS. Further, we find a Wigner crystal phase for intermediate $a_B$ and $\nu$ (\upperRomannumeral{1}), a bubble crystal for large $a_B$ (\upperRomannumeral{2}). Note that there is evidence for a liquid of clustered particles (large $a_\mathrm{B}$ and $\nu$) in the continuum ~\cite{Grusdt_2013, Gra__2018, Cooper2005, Burrello2020}, that we do not find on the lattice (\upperRomannumeral{3}). Additionally, we find first indications of a stripe phase at small filling $\nu$ with a strong magnetic field of $\alpha = 0.5$ (see Appendix~\ref{app:Stripes}). To allow a high occupation number for clustered states, we are using a local Hilbert space cut-off of $N_\mathrm{max}=3$ (see Tab.~\ref{Tab:TLR}). In the following, we discuss the phases (\upperRomannumeral{1}-\upperRomannumeral{3}) in more detail.

The various phases will be identified by the density $\rho_{i}=\langle \hat{n}_{i}\rangle$ and the two particle correlator $g^{(2)}_{i,j}$ defined by:
\begin{equation}
g^{(2)}_{i,j} = \frac{\langle \hat{n}_{i}\hat{n}_j\rangle - \delta_{i,j}\langle \hat{n}_{i}\rangle}{\langle \hat{n}_{i}\rangle\langle \hat{n}_{j}\rangle},
\end{equation}
with $i=(x_i,y_i)$ and $j=(x_j,y_j)$. In order to obtain $g^{(2)}(r)$ we average over all possible combinations of $i$ and $j$ with distance $r$. The error is estimated using the Jackknife resampling technique.


\begin{table}[b]
\centering
  \begin{tabular}{c|c|c|c|c|c}
    System & hidden dim.& lr $\times 10^{-3}$ & samples & epochs $\times 10^3$& $d_\mathrm{local}$ \\
    \hline
     $12\times12$ & 100 & 0.5-0.05 & 200 & $<100$ & $4$
  \end{tabular}
  \caption{Hyperparameters used for the benchmark. They are defined in App.~\ref{app:DefHyper}). The number of parameters is 341008 ($\leftrightarrow H_\mathrm{dim}\lesssim 10^{26}$). We used up to three different seeds: 1234, 4444, and 9999 for the calculations.}
  \label{Tab:TLR}
\end{table}

\begin{figure}
    \centering
    \includegraphics{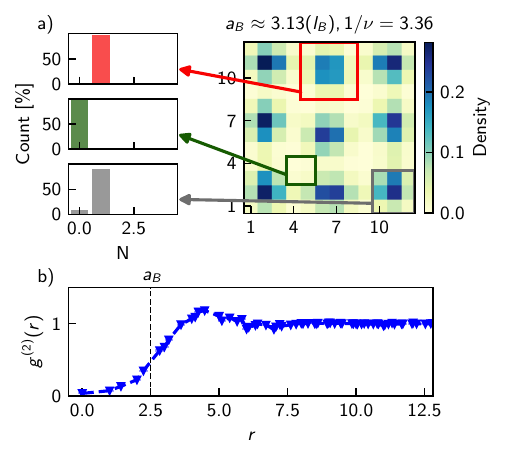}
    \caption{Density profile of a Wigner crystal with the full counting statistics of particle numbers in each square (panel a)). Panel b) shows the two particle correlator $g^{(2)}(r)$. The parameters are: $N=9$, $\alpha=0.25$, $a_\mathrm{B}=2.5$}
    \label{WignerCrystal}
\end{figure}

(\upperRomannumeral{1}) Wigner crystal: In this phase, the NQS finds a Wigner crystal (see Fig.~\ref{WignerCrystal}). The competition of long-range interactions with kinetic terms in the Hamiltonian favors localization of particles in a regular pattern. Thus, the density displayed in Fig.~\ref{WignerCrystal} shows a crystalline structure with nine crystal lattice sites for nine particles with a periodicity of $4-5$ in units of the lattice constant. This periodicity is also visible as a peak in the $g^2(r)$ correlation function, see Fig.~\ref{WignerCrystal} b). Notice that for $r\rightarrow 0$ the correlation function approaches zero, indicating that there is a maximum of one particle per site. Furthermore, the full counting statistics (FCS) of the particles in the boxes, top left in Fig.~\ref{WignerCrystal}, show a peak at one particle per crystal lattice site, and zero otherwise. These are key indications for a Wigner crystal.

When the blockade radius or the filling fraction is increased to such an extent that interactions between particles cannot be avoided, it will become favorable to form clusters of particles. Similar to the competition between the Wigner crystal and the FQH regime, we expect a competition between crystallized states (\upperRomannumeral{2}) and liquid states (\upperRomannumeral{3}). 

(\upperRomannumeral{2}) Bubble crystal:
\begin{figure}
    \centering
    \includegraphics{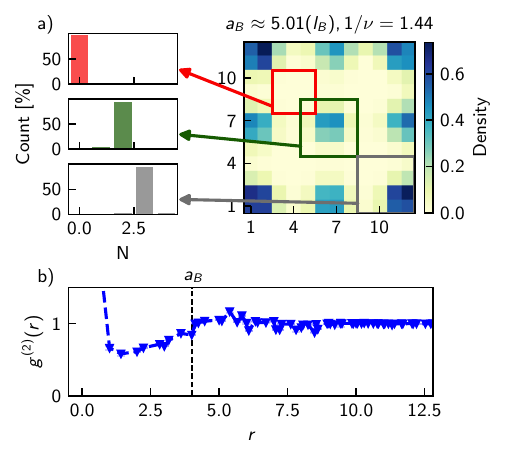}
    \caption{Density profile of a bubble crystal with the full counting statistics of particle numbers in each square ((panel a)). Panel b) shows the two particle correlator $g^{(2)}(r)$. The parameters are: $N=21$, $\alpha=0.25$, $a_\mathrm{B}=4$}
    \label{BubbleCrystal}
\end{figure}
Similar to the Wigner crystal phase, we observe a periodic lattice structure caused by the large blockade radius that hinders the movement of bubbles (see Fig.~\ref{BubbleCrystal} a)). For the parameters used in Fig.~\ref{BubbleCrystal}, we observe a crystal that has nine lattice sites, but twenty-one particles in total. Thus, the density indicates a clustering of particles. From $g^2(r)$, we obtain similar information, see Fig.~\ref{BubbleCrystal} b). The periodicity is approximately $5-6$ in units of the system's lattice constant, and $g^2(r\rightarrow0) \neq 0$ shows that there are often multiple particles close together. The FCS of the particle number in the boxes indicated in Fig.~\ref{BubbleCrystal} a) again supports this conclusion and reveals that there are three sites with bubbles of three particles each and six sites with two particles. The space in between is empty. Such a crystalline structure of clustered particles is what we refer to as bubble crystals. Notice, that the density in Fig.~\ref{BubbleCrystal} a) cannot be periodically extended without requiring three additional empty sites in the x and y direction.

Note that the boundary conditions stabilize the crystalline phases. Nonetheless, there is evidence of those symmetry broken phases in systems with no boundary effects. This is addressed in the Appendix \ref{app:SymmBroken}.

(\upperRomannumeral{3}) Cluster liquid phase: 
We do not find evidence for a cluster liquid phase on a lattice, which was theoretically predicted in the continuum at large $a_\mathrm{B}$ and $\nu$~\cite{Grusdt_2013}. This can have numerous reasons:

(i) It is likely that there is indeed no cluster liquid for the considered finite-size lattice system, as the predefined lattice of the system in combination with the large blockade radius may make free movement of clusters unfavorable. Further, we observe that the crystal lattices of bubble crystals, which compete with cluster liquids, are incommensurable with the system size. E.g. the density profile in Fig.~\ref{BubbleCrystal} cannot be used as a building block for larger systems, but would require padding of three empty sites in the x and y direction. Hence, bubble crystals lower their energy in finite-size systems by increasing the effective system size ($\nu_\mathrm{eff}<\nu$). This amounts to strong finite-size corrections at the considered system sizes. Note that its not a rare phenomenon that the incommensurability of the filling significantly impacts the physics \cite{Ng_2021, madhusudan2024, Mishra_2009, Niyaz_1994, Kuehner_1998, Batrouni_2014, Batrouni_2006}.

(ii) Although we have covered a large part of the phase diagram, it is, nonetheless, possible that we did not search in the correct parameter regime.

(iii) We know from Sec.~\ref{Benchmark12x12} that it is challenging to capture topological phases. Since cluster liquids are topologically ordered. They may be challenging to represent by our NQS.

\section{Conclusion and outlook}
In summary, we investigated the possibility to simulate FQH-like systems with RNN-NQS, extending existing numerical schemes by allowing a variable local Hilbert space dimension. To show this, we benchmarked the NQS for systems on a $6 \times 6$ and a $12 \times 12$ lattice. In the smaller system, the NQS accurately represents most of the ground state properties that we analyzed with a relative energy error of max$(\epsilon) < 0.5\%$, for all values of $\alpha$. Even in regimes where we expect to observe topological ordered phases, such as the \nicefrac{1}{2} Laughlin or the Pfaffian, we see a substantial ground state overlap of at least $80 \%$. 
Despite the fixed number of variational parameters, the evaluated observables for the much larger system were in similarly good agreement for high and low magnetic field. This indicates very good scalability, as opposed to MPS for large two-dimensional systems.  
For intermediate values of the flux per plaquette, where topological phases emerge, the obtained energies are close to the MPS value ($\epsilon \approx 1\%$). 
However, there is a significant deviation in the obtained expectation value of the Hubbard interaction. 
Therefore, we conclude that the RNN-NQS generally achieves a good approximation of the ground state. However, capturing topological order proves to be more challenging (see Appendix~\ref{app:topoligicalOrder}).

Furthermore, we showed that the RNN can be used to map out the phase diagram of the Hofstadter model with long-range interactions. Here, we find, in addition to the familiar HBH-regime, symmetry broken phases, like a Wigner crystal and a bubble crystal. We do not find evidence for the formation of a cluster liquid, which was predicted earlier in continuum systems. This may be due to strong finite-size effects even in our large lattices. We highlight that the existence of bubble crystals in an experimentally realistic model provides a promising starting point for the search for cluster liquid states that might exhibit non-Abelian braiding statistics.

In the models studied here, the NQS gives a good approximation of the ground state and is thus a unique tool for systems that have long-range interactions or require a high local Hilbert space dimension. Therefore, NQS-based methods can be particularly useful for other bosonic systems that are challenging to study with state-of-the-art numerics~\cite{BHLadder_even_2022, Saito_2018, McBrian_2019, Vargas_Calder_n_2020,denis2024accurate}. In summary, the NQS perfectly complements other numerical methods that are typically specialized tools for short-range interactions, while being able to allow high local occupations. 

Exploring the potential of other NQS architectures to capture topological order is a natural next step. Furthermore, more extensive numerical studies of similar models might be able to shed light on the exact details of the phase diagram which we sketched here. Understanding the competition of the various length scales in the model provides a promising playground for the improvement of existing and the development of new numerical methods.

\textit{Note added.--} While finishing the manuscript, we became aware of a related work by Ledinauskas and Anisimovas~\cite{ledinauskas2024universal}, studying a hard-core bosonic Hofstadter Hubbard model using a multi-layer perceptron and convolutional neural networks. Similar to our work, they find that  strong magnetic flux complicates the simulations and can lead to significant energy errors, even in numerically favorable settings.

\section*{Code and data availability}
Code and data are available from the corresponding author on request.

\begin{acknowledgments}
We thank N. K\"aming, A. Van de Walle, A. B\"ohler, T. Blatz, and R. Wiersema for fruitful discussions. We acknowledge the funding by the Deutsche Forschungsgemeinschaft (DFG, German Research Foundation) under Germany’s Excellence
Strategy – EXC-2111 – 390814868 (Grant Agreement no 948141)
and the support by the Deutsche Forschungsgemeinschaft (DFG, German Research Foundation) via Research Unit FOR 2414 under project number 277974659. This project has received funding from the European Research Council (ERC) under the European Union’s Horizon 2020 research and innovation programm (Grant Agreement no 948141) — ERC Starting Grant SimUcQuam. F.P. acknowledges support by the ERC Consolidator Grant LATIS, and the EOS Project CHEQS. H.L. acknowledges support by the International Max Planck Research School.
\end{acknowledgments}

\appendix
\section{Details for the 2D tensorized gated recurrent neural network}
\label{app:NQS}
Here, we discuss the NQS parametrization used in this work. We start with the parametrization of the wave function for the RNN-NQS, its extensions to 2D systems, and the implementation of tensorized gated RNN cells. Furthermore, we discuss the details of the parameter updates, the variational calculation of expectation values, and the particle conservation.

\subsection{Network architecture}
\subsubsection{Parametrization of the wave function}
The wave function, defined by equation \eqref{Eq1} is parametrized by an RNN with two separate output layers $P_\lambda(n)$ and $\phi_\lambda(n)$:
\begin{equation}
\begin{split}
    P_\lambda(n)=&\prod_i P_\lambda(n_i|n_{<i}) = \prod_i S_{\mathrm{sm}}(U_1 h_i +b_1)n_i,\\
    \phi_\lambda(n) =&\sum_i \phi_\lambda(n_i|n_{<i})= \sum_i S_{\mathrm{ss}}(U_2 h_i +b_2)n_i,
\end{split}
\end{equation}
where $n_i$($h_i$) is the one hot encoded particle number (hidden state) at site $i$, $S_\mathrm{sm}$ is the softmax activation function and $S_\mathrm{ss}$ is the soft sign activation function.

\subsubsection{2D RNN}
\label{app:Snakepath}
\begin{figure}
\centering
\includegraphics[scale=1.5]{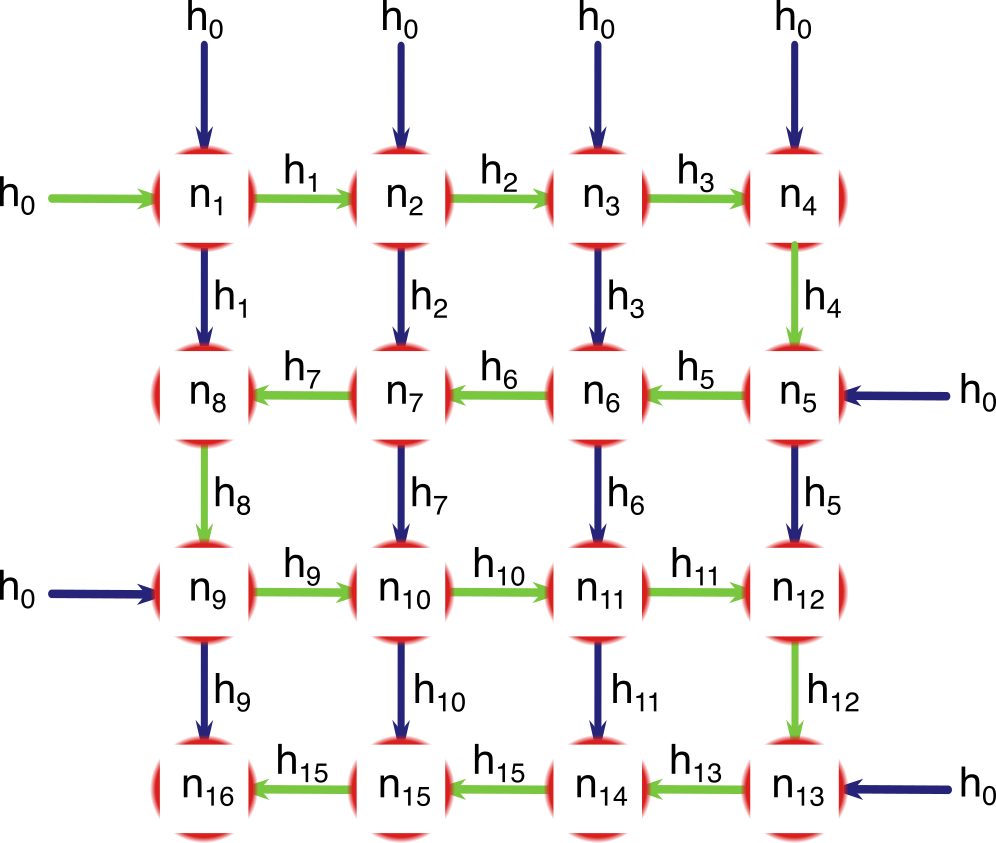}
\caption{The working principle of the machine learning algorithm. For a one-dimensional RNN-NQS, the working path and the connection of the grid sites via hidden states are the same. This is shown in green. For a two-dimensional RNN, one keeps the 1D work path but gets two hidden states (x- and y-direction). The additional sharing of hidden states is marked in blue.}
\label{app:snakePic}
\end{figure}
When solving two-dimensional systems with standard 1D RNNs, a snake path is put through the grid to calculate the system's wave function. This is illustrated in Figure~\ref{app:snakePic}. Thus, the 2D structure of the lattice is neglected and some neighboring sites of the physical system are separated in the network's path. To solve this issue, a two-dimensional RNN was proposed in~\cite{graves2007multidimensional} and later used for NQS~\cite{Hibat_Allah_2020, VariationalHibat, lange2023neural}. This RNN structure requires two inputs and two hidden states from previous sites. The details are shown in the following subsection.

\subsubsection{2D tensorized gated recurrent neural network}
\label{app:2D_RNN}
During our testing phase, the 2D tensorized gated recurrent neural network for NQS proposed in Ref.~\cite{VariationalHibat} demonstrated the best capabilities in describing ground states of two-dimensional FQH systems. The cell uses one update gate and no reset gate~\cite{zhou2016minimal}. The Network has the following structure:
\begin{align}
&u_{i,j} = \sigma(\{n^T_{i-1,j};n^T_{i,j-1}\}T_g\{h_{i-1,j};h_{i,j-1}\} + b_g),\\
&\tilde{h}_{i,j} = f(\{n^T_{i-1,j};n^T_{i,j-1}\}T\{h_{i-1,j};h_{i,j-1}\} + b),\\
&h_{i,j} = u_{i,j}\odot\tilde{h}_{i,j} + (1-u_{i,j})\odot(W\{h_{i-1,j};h_{i,j-1}\}),
\label{hidState}
\end{align}
where $T_g$ and $T$ are tensors, $i$ and $j$ denote the coordinates in the system with $h_{i,j}$ ($n_{i,j}$) being the corresponding hidden state (particle number). The activation function $f$ is a hyperbolic tangent but can be exchanged if other options are preferable. In particular, the update gate $u_{i,j}$ and the two-dimensional structure of the network are crucial for reliable approximations~\cite{Hibat_Allah_2020}. 

\subsection{Update procedure and optimizer}
\label{AppUpdate}
To find the ground state during the training procedure, the network parameters need to be updated such that the energy is minimized. To do so, one of the two variational methods is typically used:
\begin{itemize}
\item[1)] Update according to the derivative of the energy~\cite{Carleo_2017Solving}
\item[2)] Update according to the natural gradient of the quantum state (stochastic reconfiguration)\cite{Sorella1998}
\end{itemize}
We use method 1), which directly uses the derivative of the energy for the update:
\begin{equation}
\label{App:loss}
\frac{\partial \langle \hat{H} \rangle}{\partial w} \approx 2 \Re (\langle O^*_w(n) H \rangle - \langle O^*_w(n)\rangle \langle H \rangle ),
\end{equation}
with $O_w(n) = \frac{1}{\psi_\lambda(n)} \frac{\partial\psi_\lambda (n)}{\partial w} =\frac{\partial}{\partial w} \log(\psi_\lambda (n))$. The parameterss are then optimized according to~\cite{Carleo_2017Solving}: $w \rightarrow w - \gamma \frac{\partial \langle \hat{H} \rangle}{\partial w}$.

Typical optimizers for optimization are stochastic gradient descent, Adam~\cite{kingma2017adam}, or AdaBound~\cite{luo2019adaptive}. In this research we use AdaBound as we empirically found it to be the optimizer with the most constant good results (see Appendix~\ref{comp6x6Sec}). The AdaBound optimizer aims to provide a similarly fast and but more generalized result, by initially using the speed of the Adam optimizer while slowly switching to the SGD with momentum. To achieve this, the adaptive learning rate from Adam is clipped by a moving lower and upper boundary $[\eta_l,\eta_u]$. The boundary functions are defined as: $\eta_l = \mathrm{lr}_f * (1 - 1 / (10^{-3} \times \mathrm{epoch} + 1))$ and $\eta_u = \mathrm{lr}_f * (1 + 1 / (10^{-3} \times \mathrm{epoch}))$, where $\mathrm{lr}_f$ is the final learning rate. For a detailed description of this method, we refer to Ref.~\cite{luo2019adaptive}.

Method 2) is often a good choice for the ground state search since it incorporates the knowledge of the geometrical structure of the optimization landscape. However, due to the immense number of parameters used in the tensorized gated RNN, stochastic reconfiguration becomes intractable. Methods like minSR allow an SR training even for deep networks with many parameters~\cite{chen2023efficient}, but empirically, we found that this method struggles with RNNs. A more detailed analysis with a similar conclusion can be found in~\cite{lange2023neural, Donatella2023}.

\subsubsection{Comparing different optimizers}
\label{comp6x6Sec}
\begin{figure}
\centering
\includegraphics{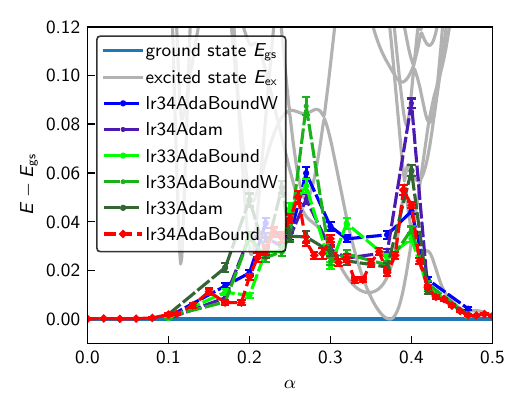}
\caption{Comparison of the optimizers: Adam, AdaBound, AdaBoundW with a static initial learning rate of $0.5\times10^{-3}$ (lr33) or a  decay to $0.5\times10^{-4}$ (lr34). Used configuration: Hofstadter-Bose-Hubbard model $6\times6$, $N=4$, $N_\mathrm{max}=2$, $U=4t$. All displayed configurations were trained for 200000 epochs.}
\label{comp6x6}
\end{figure}

One of the most time-consuming problems for machine learning tasks is the search for the best hyperparameters and optimizers. During our optimization phase, we tested many different optimizers with multiple learning rates. In Figure~\ref{comp6x6} we compare the Adam~\cite{kingma2017adam}, AdaBound~\cite{luo2019adaptive}, and AdaBound with decoupled weight decay~\cite{loshchilov2019decoupled}. The best working learning rates for all optimizers were of the order $10^{-3}$. In all cases we used $0.5\times10^{-3}$ as starting learning rate, that stayed constant (lr33) or decayed gradually to $0.5\times10^{-4}$ (lr34) during the training. The decay is given by $\mathrm{lr}=\mathrm{max}(\mathrm{lr_{threshold}},\mathrm{lr}_\mathrm{initial}/(1+\mathrm{epoch}/500))$. The AdaBound optimizer switches to a final learning rate $\mathrm{lr}_f$ of $0.01$ for lr34 and to $0.1$ for lr33.
Mostly, we find that the AdaBound optimizer with lr34 is one of the best choices, although there are some exceptions where the lr34AdaBound is outperformed.

We additionally tested the Adam optimizer for a learning rate of lr22, lr44, and lr55. The first one led to exploding gradients, the other two had worse results than the results presented for lr33 and lr34. 

\subsubsection{Additional tools for the ground state search}
Besides the optimization methods introduced above, other approaches for the ground state search were tested. One method in combination with the RNN-NQS that was shown to be advantageous for some spin systems is variational classical annealing~\cite{AnnealingHibat_Allah, VariationalHibat}, where an entropy term is added to the loss function. This simulates a non-zero temperature to achieve an easier exploration of the Hilbert space at a high artificial temperature. However, we did not see significant improvements when applied to this model.
Furthermore, one can directly enforce spatial symmetries with the RNN-NQS~\cite{Hibat_Allah_2020,Reh_2023}, which is, however, not possible for the FQH effect, due to its non-symmetric phase structure. Lastly, the RNN architecture can be used with and without weight sharing. In most cases, the RNN is used with weight sharing, explicitly exploiting the translational invariance of the system under consideration. For systems without translational invariance, RNNs without weight sharing, i.e. different network parameters for each lattice site, can improve the performance of the ansatz~\cite{VariationalHibat}. We tested this ansatz using a site-dependent output layer for the phase, to account for the non-symmetric phase structure of FQH states. This did not lead to an improvement of our results.

\subsection{Variational evaluation of observables}
\label{AppVMC}
The expectation value of an observable of a system in state $|\psi\rangle= \sum_n \psi(n)|n\rangle$ is defined by:
\begin{equation}
\langle \hat{A} \rangle \equiv \frac{\langle \psi| \hat{A} |\psi\rangle}{\langle\psi|\psi\rangle} = \sum_{n,n'}\langle n|\hat{A}|n' \rangle \frac{\psi(n')}{\psi(n)}\frac{|\psi(n)|^2}{\sum_n |\psi(n)|^2},
\label{AppVMC:Eq1}
\end{equation}
where $|n\rangle$ is sampled from the probability distribution $|\psi(n)|^2$. Note that $\langle\psi|\psi\rangle = 1$ is already normalized due to the autoregressive parametrization of the wave function. Instead of summing over all possible states $n$, one can approximate equation \eqref{AppVMC:Eq1} by summing over all sampled states:
\begin{equation}
  \langle \hat{A} \rangle \approx \frac{1}{N_\text{s}} \sum_{n_\text{s} \sim |\psi(n)|^2} \underbrace{\sum_{n'} \langle n_\text{s}|\hat{A}|n' \rangle \frac{\psi(n')}{\psi(n_\text{s})}}_{A_\mathrm{loc}(n_\text{s})}.
\label{AppVMC:Eq2}
\end{equation}

\subsection{\texorpdfstring{ $U(1)$}{TEXT}-symmetry: particle conservation}
\label{AppU1}
A crucial detail for NQS calculations is the control over the particle $N_\mathrm{RNN}$. In principle, it is possible to do the calculations grand canonically, by regulating a chemical potential. However, it is often advantageous to use a fixed and conserved number of particles. The algorithm has the following structure:
\begin{enumerate}
\item{Check if the local Hilbert space cut-off $N_\mathrm{max}$ and the total number of particles $N_{\mathrm{total}}$ allow further unrestricted placing of particles.}
\item{Set maximum particle number $n^{j}_{\mathrm{max}, i} \leq N_\mathrm{max}$ for a given sample $j$ at site $i$, to prevent snapshots with too many particles: $N_\mathrm{RNN}>N_\mathrm{total}$.}
\item{Set minimum particle number $n^{j}_{\mathrm{min}, i} \geq 0$ for a given sample $j$ at site $i$, to prevent too few particles per snapshot: $N_\mathrm{RNN}<N_\mathrm{total}$.}
\item{Set the probability of all prohibited choices in the vector $S_\mathrm{sm}(U_1 h_i +b_1)$ to zero. Renormalize the probability vector $S_\mathrm{sm}(U_1 h_i +b_1)$.}
\end{enumerate}

\subsection{Hyperparameters}
\label{app:DefHyper}
The RNN-NQS can be finetuned by varying multiple hyperparameters. Here, we define the parameters that we actively tune. To control the total number of parameters we use the hidden dimension $d_h$ (dimension of the hidden vector in Eq.~\ref{hidState}). The total number of parameters is given by $n_p = (8 d_\mathrm{local} + 2)d_h^2 + 2(d_\mathrm{local}  +1)d_h + 2 d_\mathrm{local}$, with the local Hilbert space dimension $d_\mathrm{local}$. To update the parameters we use the AdaBound optimizer (see App.~\ref{AppUpdate}) with an initial learning rate of $\mathrm{lr}_\mathrm{initial}=0.0005$ that decays to $\mathrm{lr}_\mathrm{initial}=0.00005$. The decay is given by $\mathrm{lr}_\mathrm{initial}/(1 + \mathrm{epoch}/500)$. Note that the AdaBound optimizer uses upper and lower boundaries for the learning rate that converge to $lr_f = 0.01$ (for a detailed description we refer to App.~\ref{AppUpdate} and Ref.~\cite{luo2019adaptive}). For each training step (epoch) we use 200 samples to evaluate the loss function, defined in Eq.~\ref{App:loss}.

\section{Details of the DMRG simulations}
\label{app:DMRG}
For systems of $L_x\times L_y = 12\times 12$ sites, we use density-matrix renormalization-group simulations~\cite{White1992,Schollw_ck_2011}, using the \textsc{SyTen}-toolkit~\cite{HubigSyTen}.
We used the single-site variant of the DMRG method with subspace expansion~\cite{Hubig2015} and exploited the $\mathrm{U}(1)$-symmetry associated with particle-number conservation.
We truncated the local Hilbert space to at most $N_{\rm max}=2$ bosons per site and used bond dimensions up to $\chi=2048$.
This results in a number of variational parameters on the order $\mathcal{O}\left((N_{\rm max}+1) \times N_{\rm sites} \times \chi^2\right) = \mathcal{O}\left(10^{9}\right)$, which is significantly larger than the number of parameters used for the NQS ($260.806$).
To reach a number of variational parameters comparable to those of the NQS, we also performed MPS simulations at a maximum bond dimension of $\chi=32$.

\begin{figure}
\centering
\includegraphics{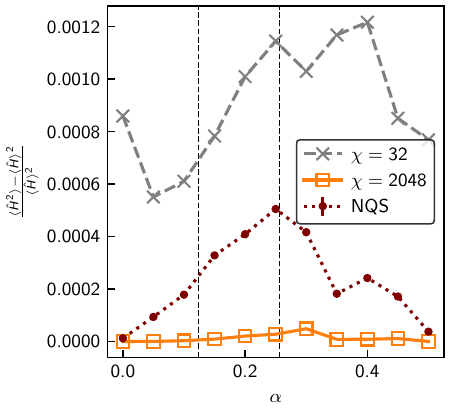}
\caption{The relative error a) and the normalized variance $(\braket{\hat{H}^2} - \braket{\hat{H}}^2)/\braket{\hat{H}}^2$ b) obtained from the NQS and the DMRG simulations at bond dimension $\chi=32$ and $2048$ .
}
\label{12x12_DMRG_Conv}
\end{figure}

We find that the normalized variance of the energy, $\frac{\braket{\hat{H}^2} - \braket{\hat{H}}^2}{\braket{\hat{H}}^2}$, is on the order of $10^{-3}$ for $\chi=32$, while it is one order of magnitude smaller for $\chi=2048$, see Fig.~\ref{12x12_DMRG_Conv}.
This accuracy is sufficient for a comparison with the NQS simulations in the main text.

\section{Additional details on the benchmark}
\label{app:Benchmark}
\begin{figure}
    \centering
    \includegraphics{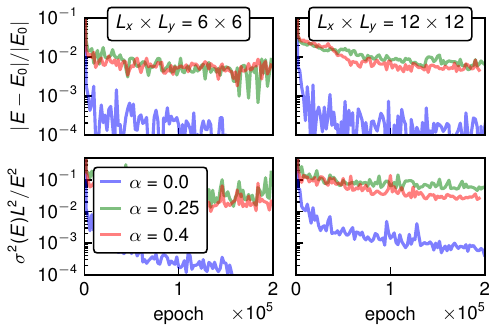}
    \caption{A comparison of exemplary training procedures for a Hofstadter model on a $6 \times 6$ ($N=4$ particles) and $12 \times 12 $ ($N=16$ particles) square lattice with on-site interactions. This figure shows the relative error in the upper and the normalized statistical variance of the energy in the lower picture.}
    \label{12x12Conv}
\end{figure}

To obtain additional information about our results, we compare the relative error and the normalized variances of the energy. The normalized quantum mechanical variance of the $12 \times 12$ benchmark can be seen in Fig.~\ref{12x12_DMRG_Conv}. Here, we observe that the variance in the regime (\upperRomannumeral{1}) ($\alpha = 0.1$ and $\alpha \gtrsim 0.35$) is smaller than for the intermediate regime (\upperRomannumeral{2}). In Fig.~\ref{12x12Conv}, selected points from the $6 \times 6$ and $12 \times 12$ benchmarks are shown. Here, we show the relative error and the normalized statistical variance as a function of the epoch. Note that the statistical variance is evaluated with PyTorch's internal function that calculates: $\sigma^2(E)=(\langle \hat{H}^2 \rangle - |\langle \hat{H} \rangle|^2)$.
Interestingly, the graphs for the $6\times6$ and for the $12\times 12$ lattice are similar in their structure and convergence speed for 200000 epochs. For $\alpha = 0$, both systems converge comparatively fast to the ground state. The other cases take significantly longer to get to the ground state. Note that the variance, for both system sizes, is lower for $\alpha = 0.4$ than for $\alpha = 0.25$, although the error is approximately the same. On the $6\times 6$ lattice, the lower variance of the NQS for $\alpha = 0.4$ can be explained by the, on average, smaller energetic differences of the contributing ground and first/second excited state (see Fig. \ref{FigHofOverlap}). The NQS for $\alpha = 0.25$ ($6\times 6$) has a much larger ground state contribution but a higher variance due to the significantly larger energy difference between ground state and first excited state.

\section{Additional Data}
\label{app:AdditionalData}
\subsection{Representational power of RNN-NQS}
\begin{figure}
    \centering
    \includegraphics{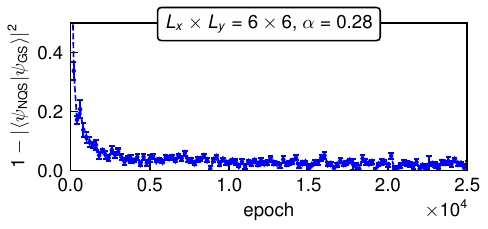}
    \caption{Fidelity optimization. We show the infidelity as a function of epochs for the interacting Hofstadter model on a $6 \times 6$ lattice. The parameters are: $N = 4$, $\alpha = 0.28$, $N_\mathrm{max}=2$, $U =4$.}
    \label{app:6x6Infid}
\end{figure}

In the benchmark section \ref{BenchmarkHubbardInteractions}, we observed that the NQS did not always capture the full ground state (compare to Figure \ref{FigHofOverlap}). To show evidence that the NQS did not reach its representational limit, we show an exemplary fidelity optimization of the $6 \times 6$ system with a magnetic flux of $\alpha = 0.28$ in Figure \ref{app:6x6Infid}. Already after a few thousand epochs, the ground state overlap surpasses the measured overlap of the NQS with the minimization according to the energy gradient ($\sim 89.7\%$ after 200K epochs). We stopped the training after 25K epochs, where we find an ground state overlap of approximately $98\%$. Based on these findings we expect a perfect ground state overlap after further training. 
Hence, we find clear evidence that the optimization according to the energy gradient is the limiting factor, while the representational limit of the RNN-NQS is not reached for the shown results in section \ref{BenchmarkHubbardInteractions}.

\subsection{Topologically ordered states}
\label{app:topoligicalOrder}
In this section, we address the challenges of finding topological ordered states. To avoid finite-size effects and distortions from the boundary, we work with periodic boundary conditions (PBC). The considered topologically ordered state is at $\nu = \nicefrac{1}{2}$ filling. The hyperparameters are given in Tab.~\ref{appTab:TB}.
\begin{table}[b]
\centering
  \begin{tabular}{c|c|c|c|c|c}
    System & hidden dim.& lr $\times 10^{-3}$ & samples & epochs $\times 10^3$& $d_\mathrm{local}$  \\
    \hline
     $6\times6$ & 100 & 0.5 & 200 & $\leq 600$ & $4$
  \end{tabular}
  \caption{Hyperparameters for the discussion of topologically ordered states. The parameters are defined in App.~\ref{app:DefHyper}. We used the seed 1234.}
  \label{appTab:TB}
\end{table}

As discussed in section~\ref{BenchmarkHubbardInteractions}, the ground state in topologically ordered phases is harder to represent for the RNN-NQS than for topological trivial phases. We observe correctly captured topological states for the $6\times6$ lattice, but complications for the $12\times 12$ lattice. Since boundary effects play a major role on the $6 \times 6$ lattice, we assume that these can help to capture the topological ordered ground state. In the following, we discuss results from a $6\times 6$ lattice with 3 particles and a magnetic flux per plaquette of $\alpha = 1/6$ (PBC). The implemented Hamiltonian is given in Eq. \eqref{HBHeq}. The Hubbard interaction is $U/t=1$.

\begin{figure}
\centering
\includegraphics{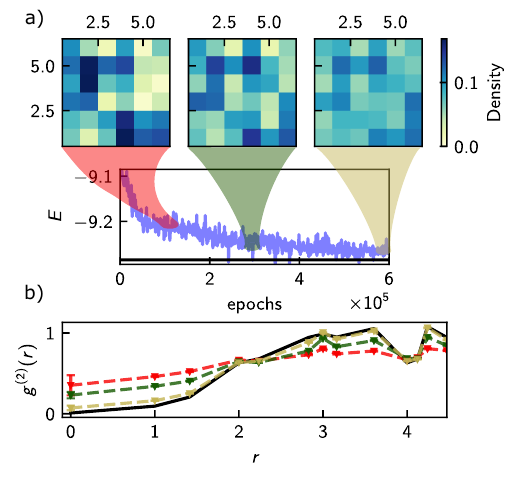}
\caption{Improvement over the epochs. Here, we compare three different training steps: red = 100K epochs, green = 300K epochs, and yellow = 600K. Panel a) displays energy convergence with densities corresponding to the three training steps. Panel b) shows the two-particle correlator at the corresponding training steps. The black line display the exact results for the ground state energy and the two-particle correlator. The Hamiltonian parameters for this $6 \times 6$ lattice are:  $N = 3$, $\alpha = 1/6$, $N_\mathrm{max}= 3$, $U = 1$.}
\label{6x6PBC}
\end{figure}

In Fig.~\ref{6x6PBC}, we compare the NQS output at different training steps. At 100K epochs, we find an inhomogeneous state, that has an energy close to the ground state energy, but the two-particle correlator deviates from the exact result (similar to deviations in the $C^{(2)}$ values for $12 \times 12$ OBC lattice). This improves by further training. At 300K epochs, the previous maxima and minima are smeared out, such that the density is more homogeneous. The energy is closer to the ground state energy and the $g^{(2)}$ value has improved.  At 600K epochs, the NQS energy deviates only slightly from the ED result and the improved NQS state has a density, that resembles a homogeneous Laughlin state. Also, the two-particle correlator is in good agreement with the exact result. We assume that the NQS will perfectly capture the ground state with further training.

This shows that the NQS can capture the topological ordered Laughlin state at least for small system sizes. Nonetheless, success depends heavily on the number of epochs. Also, the effect of boundaries indeed seems to play a role, as the NQS takes significantly longer to converge to the ground state compared to the benchmarked $6 \times 6$ open boundary case (\ref{Benchmark6x6}). Note that the Hilbert space dimension of the OBC cases is, due to the difference in the number of particles, about 8 times larger compared to this PBC case.
Thus, an even larger system, where boundary effects only play a minor role, will most likely require additional computational resources. This explains why topological states on the $12 \times 12$ lattice (OBC, see section~\ref{Benchmark12x12}), were not properly captured after 200K epochs.

\subsection{Stripe phase}
\label{app:Stripes}
\begin{figure}
    \centering
    \includegraphics{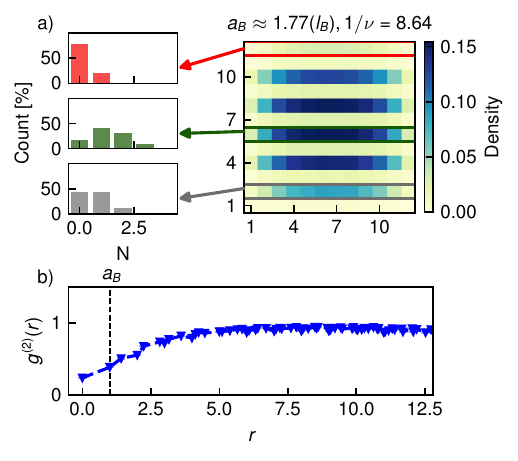}
    \caption{Density profile of a typical stripe order with the full counting statistics of particle numbers in each rectangle (panel a)). Panel b) shows the two particle correlator $g^{(2)}(r)$. The parameters are: $N=7$, $\alpha=0.5$, $a_\mathrm{B}=1.0$}
    \label{Stripes}
\end{figure}

During our study of long-range systems, we also found a stripe phase for a large magnetic field and low filling. Fig.~\ref{Stripes} is an example of this, where the density has vacancies in every second row. Although there are in total 6 stripes for seven particles, we expect no clustering within the stripes, as $g^2(r)$ is comparatively small for $r\rightarrow 0$, see Fig.~\ref{Stripes} b). The overall $g^2(r)$ structure is similar to a liquid since there is no peak at some distance $r$. This observation extends into the stripes, as we can conclude from the widely distributed particle number per stripe in the histograms. Thus, we have a liquid behavior in one direction while having a spatial periodic structure in the other. 

Note that in the context of long-range interactions in a magnetic field, other studies also observed stripe phases in the continuum~\cite{Gra__2018, Cooper2005, Burrello2020}. However, these were found close to $\nu = \nicefrac{1}{2}$ at a much smaller magnetic field, which is why direct connection between those phases cannot be readily drawn.

\subsection{Symmetry broken phases in systems without stabilizing boundary effects}
\label{app:SymmBroken}
\begin{figure}
    \centering
    \includegraphics{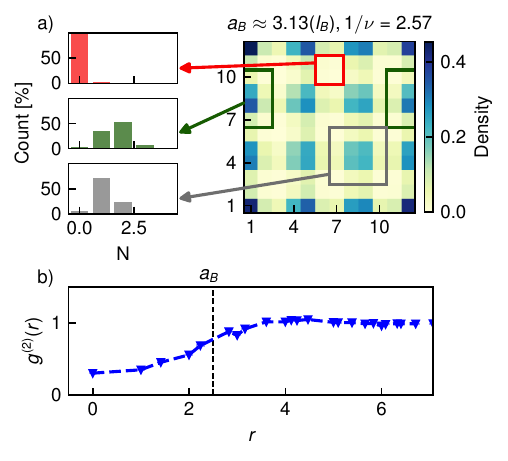}
    \caption{Density profile and the full counting statistics of particle numbers in each square (panel a)) of a crystalline structure in a system with periodic boundaries. Panel b) shows the two particle correlator $g^{(2)}(r)$. The parameters are: $N=14$, $\alpha=0.25$, $a_\mathrm{B}=1.0$}
    \label{BubbleCrystal_PBC}
\end{figure}
In the subsection \ref{phaseD}, we identified two crystalline phases for the considered long-range interacting system. However, these symmetry broken phases often lower their energy by increasing the effective system size, which can make them favorable compared to liquid-like phases.

Nonetheless, there is evidence that the symmetry broken phases persist in systems where the crystalline structure is not stabilize by the boundary conditions. In Figure \ref{BubbleCrystal_PBC}, we show a crystalline phase in a system with periodic boundary conditions. The respective density profile shows 9 crystal sites for 14 particles. Combined with the FCS that indicates $1-2$ particles per crystal lattice site, we can infer that there are on average 5 crystal sites double occupied. The crystal sites have a periodicity of $4$, which is also supported by the two particle correlator that shows a small maximum at $r= 4$.

The evidence of crystalline phases in systems with periodic boundary conditions, indicates that we can expect symmetry broken phases in systems with less stabilizing boundary effects.

Note that using PBCs the ground state might be a superposition of crystalline configurations related by translations. However, due to our implementation of the sampling, our simulations favor variational ground states with a high occupation at the first few lattice sites. Nonetheless, there is remaining evidence of a potential superposition, since the full counting statistic is less peaked around a certain particle number.

\bibliographystyle{apsrev4-2}
\bibliography{biblio}

\end{document}